\documentclass[twocolumn,aps,prd,superscriptaddress,nofootinbib,floatfix,preprintnumbers]{revtex4-1}

\usepackage{bm,amsmath,amsfonts,amssymb,graphicx,xspace,placeins,multirow}

\usepackage{hyperref}
\usepackage{breakurl}
\usepackage{array}
\usepackage{slashed}
\usepackage{xspace}

\allowdisplaybreaks
\newcommand{\nn}{\nonumber}

\def\d{{\rm d}}

\newcommand{\g}{\gamma}

\newcommand{\cbar}{\bar{c}}
\newcommand{\Bbar}{\,\overline{\!B}{}}

\newcommand{\lqcd}{\ensuremath{\Lambda_{\rm QCD}}\xspace}

\newcommand{\ampBb}[2]{\big\langle #1 \big|\, #2\, \big| \Bbar \big\rangle}
\renewcommand{\ampBb}[2]{\langle #1 |\, #2\, | \Bbar \rangle}

\newcommand{\rDs}{r_{D^*}}

\newcommand{\clnp}{CLNnoR\xspace}
\newcommand{\nocln}{noHQS\xspace}

\makeatletter
\g@addto@macro\bfseries{\boldmath}
\makeatother

\usepackage[dvipsnames]{xcolor}
\definecolor{red}{rgb}{0.9, 0,0}

\begin{document}

\title{Tensions and correlations in $|V_{cb}|$ determinations}

\author{Florian U.\ Bernlochner}
\affiliation{Physikalisches Institut der Rheinischen Friedrich-Wilhelms-Universit\"at Bonn, 53115 Bonn, Germany}
\affiliation{Karlsruher Institute of Technology, 76131 Karlsruhe, Germany}

\author{Zoltan Ligeti}
\affiliation{Ernest Orlando Lawrence Berkeley National Laboratory, 
University of California, Berkeley, CA 94720, USA}

\author{Michele Papucci}
\affiliation{Ernest Orlando Lawrence Berkeley National Laboratory, 
University of California, Berkeley, CA 94720, USA}

\author{Dean J.\ Robinson}
\affiliation{Physics Department, University of Cincinnati, Cincinnati OH 45221, USA}

\begin{abstract}

Recently several papers extracted $|V_{cb}|$ using the Belle
measurement~\cite{Abdesselam:2017kjf} of the exclusive $\bar B \to D^*
\ell\bar\nu$ unfolded differential decay rates, available for the first time.
Depending on the theoretical inputs, some of the fits yield higher $|V_{cb}|$
values, compatible with those from inclusive semileptonic $B$ decays.  Since
these four fits use mostly the same data, if their correlations were close to
100\%, the tension between them would be over $5\sigma$.  We determine the
correlations, find that the tension between the results is less than
$3\sigma$, and explore what might lead to improving the consistency of the
fits.  We find that fits that yield the higher values of $|V_{cb}|$, also
suggest large violations of heavy quark symmetry.  These fits are also in
tension with preliminary lattice QCD data on the form factors.  Without
additional experimental data or lattice QCD input, there are no set of
assumptions under which the tension between exclusive and inclusive
determinations of $|V_{cb}|$ can be considered resolved.

\end{abstract}

\maketitle

\section{introduction}

Using the unfolded $\bar B \to D^* \ell\, \bar\nu$
spectra from Belle~\cite{Abdesselam:2017kjf}, several theory
papers~\cite{Bernlochner:2017jka, Bigi:2017njr, Grinstein:2017nlq} could
perform fits to the data for the first time, using different theoretical
approaches.  Using the BGL parametrization~\cite{Boyd:1995sq,
Boyd:1997kz} for the $\bar B \to D^* \ell \, \bar\nu$ form factors, a
substantial shift in the extracted value of $|V_{cb}|$ was
found~\cite{Bigi:2017njr, Grinstein:2017nlq}, compared to the
Belle~\cite{Abdesselam:2017kjf} analysis using the CLN~\cite{Caprini:1997mu}
parametrization,
\begin{subequations}
\begin{eqnarray}
|V_{cb}|_{\rm CLN} &=& (38.2\pm 1.5)\times 10^{-3}\,,
  \quad\,\text{\cite{Abdesselam:2017kjf}}\,, \label{belleVcb}\\
|V_{cb}|_{\rm BGL} &=& (41.7^{+2.0}_{-2.1})\times 10^{-3}\,, 
  \qquad \text{\cite{Bigi:2017njr}}\,, \label{gambinoVcb}\\
|V_{cb}|_{\rm BGL} &=& (41.9^{+2.0}_{-1.9})\times 10^{-3}\,, 
  \qquad \text{\cite{Grinstein:2017nlq}}\,. \label{benVcb}
\end{eqnarray}
\end{subequations}
The main result in Ref.~\cite{Abdesselam:2017kjf} was $|V_{cb}|_{\rm
CLN} = (37.4\pm 1.3)\times 10^{-3}$, obtained from a fit inside the Belle
framework, before unfolding.  Only Eq.~(\ref{belleVcb}) quoted in the
Appendix of~\cite{Abdesselam:2017kjf} can be directly compared with
Eqs.~(\ref{gambinoVcb}) and (\ref{benVcb}).  These papers, as well as this
work, use the same fixed value of ${\cal F}(1)$~\cite{Bailey:2014tva} (see
Eq.~(\ref{dGdw}) below), so the differences in the extracted values of
$|V_{cb}|$ are due to the extrapolations to zero recoil, where heavy quark
symmetry gives the strongest constraint on the rate~\cite{Isgur:1989vq,
Isgur:1989ed, Shifman:1987rj, Nussinov:1986hw, Luke:1990eg}. Intriguingly,
the BGL fit results for $|V_{cb}|$ are compatible with those from inclusive
$B\to X_c\ell\bar\nu$ measurements~\cite{HFAG}.  If one assumed, naively, a
100\% correlation between the fits yielding Eqs.~(\ref{belleVcb}),
(\ref{gambinoVcb}), and (\ref{benVcb}), then the tension between
Eqs.~(\ref{belleVcb}) and (\ref{gambinoVcb}) or between Eqs.~(\ref{belleVcb})
and (\ref{benVcb}) would be above $5\sigma$.

The BGL~\cite{Boyd:1995sq, Boyd:1997kz} fit implements constraints on the
$B\to D^*\ell\bar\nu$ form factors based on analyticity and
unitarity~\cite{Bourrely:1980gp, Boyd:1994tt, Boyd:1995cf}.  The
CLN~\cite{Caprini:1997mu} fit imposes, in addition, constraints on the form
factors from heavy quark symmetry, and relies on QCD sum rule
calculations~\cite{Neubert:1992wq, Neubert:1992pn, Ligeti:1993hw} of the
subleading Isgur-Wise functions~\cite{Luke:1990eg, Falk:1990pz}, without
accounting for their uncertainties.  Ref.~\cite{Bernlochner:2017jka}
performed combined fits to $\bar B\to D^* \ell\, \bar \nu$ and $\bar B\to D
\ell\, \bar \nu$, using predictions of the heavy quark effective theory
(HQET)~\cite{Georgi:1990um, Eichten:1989zv}, including all ${\cal
O}(\lqcd/m_{c,b})$ uncertainties and their correlations for the first time. 
The effect of relaxing the QCD sum rule inputs in the CLN fit was found to be
small compared to the difference of the CLN and BGL results.

The recent papers using the BGL parametrization~\cite{Bigi:2017njr,
Grinstein:2017nlq} assert that the higher values obtained for $|V_{cb}|$ are
due to the too restrictive functional forms used in the CLN fits.  It was
previously also noticed that the CLN gives a poorer fit to the $B\to
D\ell\bar\nu$ data than BGL~\cite{Bigi:2016mdz}.  The effects on $|V_{cb}|$
due to additional theoretical inputs were also explored in
Refs.~\cite{Bigi:2017jbd, Jaiswal:2017rve}.

Based on our work in Ref.~\cite{Bernlochner:2017jka}, we explore which
differences between the BGL and CLN fits are responsible for the different
extracted $|V_{cb}|$ values, study the consistency and compatibility of the
fits, and the significance of the shift in the extracted value of
$|V_{cb}|$. 

\begin{table*}[bt]
\begin{tabular}{c|cccc}
\hline\hline
form factors  &  BGL  &  CLN  &  \clnp  &  \nocln  \\
\hline
axial $\propto \epsilon^*_\mu$
&  $b_0,\, b_1$  &  $h_{A_1}(1),\ \rho_{D^*}^2$
  &  $h_{A_1}(1),\ \rho_{D^*}^2$
  &  $h_{A_1}(1),\ \rho_{D^*}^2,\ c_{D^*}$  \\
vector
  &  $a_0,\, a_1$
  &  \multirow{2}{*}{$\bigg\{ R_1(1),\, R_2(1)$}
  &  \multirow{2}{*}{$\bigg\{$ \hspace*{-10pt} \begin{tabular}{l}
  	$R_1(1),\ R'_1(1)$ \\ $R_2(1),\ R'_2(1)$\end{tabular}}
  &  \multirow{2}{*}{$\bigg\{$ \hspace*{-10pt} \begin{tabular}{l}
  	$R_1(1),\ R'_1(1)$ \\ $R_2(1),\ R'_2(1)$\end{tabular}} \\
${\cal F}$  &  $c_1,\, c_2$  \\
\hline\hline
\end{tabular}
\caption{The fit parameters in the BGL, CLN, \clnp, and \nocln
fits, and their relationships with the form factors.}
\label{fitsummary}
\end{table*}

\section{definitions}

The $B\to D^{*}\ell\bar\nu$ form factors which occur in the standard model
are defined as
\begin{align}\label{formfactors}
\ampBb{D^*}{\cbar \g^\mu b} & = i\sqrt{m_B m_{D^*}}\, h_V\, 
  \varepsilon^{\mu\nu\alpha\beta}\,
  \epsilon^*_{\nu}v'_\alpha v_\beta \,, \nn\\*
\ampBb{D^*}{\cbar \g^\mu \g^5 b} & = \sqrt{m_B m_{D^*}}\, 
  \big[h_{A_1} (w+1)\epsilon^{*\mu} \\*
  & \quad - h_{A_2}(\epsilon^* \cdot v)v^\mu 
  - h_{A_3}(\epsilon^* \cdot v)v'^\mu \big] , \nn
\end{align}
where $v$ is the four-velocity of the $B$ and $v'$ is that of the $D^{*}$.
The form factors $h_{V,A_{1,2,3}}$ depend on $w = v\cdot v' =
(m_B^2+m_{D^{*}}^2-q^2) / (2m_B m_{D^{*}})$. Neglecting lepton masses, only
one linear combination of  $h_{A_2}$ and $h_{A_3}$ is measurable. In the
heavy quark limit, $h_{A_1} = h_{A_3} = h_V =  \xi$ and $h_{A_2} = 0$, where
$\xi$ is the Isgur-Wise function~\cite{Isgur:1989vq, Isgur:1989ed}.  Each of
these form factors can be expanded in powers of $\lqcd / m_{c,b}$ and
$\alpha_s$.  It is convenient to parametrize deviations from the heavy quark
limit via the form factor ratios
\begin{equation}\label{eqn:R1R2Def}
  R_1(w) = \frac{h_V}{h_{A_1}}\,, \qquad 
  R_2(w) = \frac{h_{A_3} + \rDs h_{A_2}}{h_{A_1}}\,,
\end{equation}
which satisfy $R_{1,2}(w) = 1 + {\cal O}(\lqcd/m_{c,b},\, \alpha_s)$ in the
$m_{c,b} \gg \lqcd$ limit, and $\rDs = m_{D^*}/m_B$.

The $B\to D^*\ell\bar\nu$ decay rate is given by
\begin{align}\label{dGdw}
\frac{\d \Gamma}{\d w} & = \frac{G_F^2|V_{cb}|^2\,  m_B^5}{48 \pi^3}\,
  (w^2-1)^{1/2}\, (w + 1)^2\, \rDs^3 (1- \rDs)^2 \nn \\*
  & \quad \times \bigg[1 + \frac{4w}{w+1}
  \frac{1- 2 w\rDs + \rDs^2}{(1 - \rDs)^2} \bigg] \mathcal{F}(w)^2\,,
\end{align}
and the expression of ${\cal F}(w)$ in terms of the form factors defined in
Eq.~(\ref{formfactors}) is standard in the literature~\cite{Manohar:2000dt}. 
In the heavy quark limit, ${\cal F}(w) = \xi(w)$.  We further denote
\begin{equation}
\rho_{D^*}^2 = -\frac1{h_{A_1}(1)} \frac{\d h_{A_1}(w)}{\d w}\bigg|_{w=1} \,,
\end{equation}
which is a physical fit parameter in the CLN approach, and is a derived
quantity in the other fits.

\section{New fits, lattice QCD, and their tensions}

The constraints built into the CLN fit can be relaxed by ignoring the QCD sum
rule inputs and the condition $R_{1,2}(w) = 1 + {\cal O}(\lqcd/m_{c,b},\,
\alpha_s)$ following from heavy quark symmetry. 
(Ref.~\cite{Bernlochner:2017jka} showed that only ignoring the QCD sum rule
inputs, and using only $w=1$ lattice QCD data, leaves $|V_{cb}| = (38.8 \pm 1.2) \times 10^{-3}$.)  Thus, we write
\begin{align}
R_1(w) & = R_1(1) + (w-1) R_1'(1) \,, \nn\\*
R_2(w) & = R_2(1) + (w-1) R_2'(1) \,,
\end{align}
and treat $R_{1,2}(1)$ and $R_{1,2}'(1)$ as fit parameters. We refer to this
fit as ``\clnp''.  It has the same number of fit parameters as BGL, and allows ${\cal
O}(1)$ heavy quark symmetry violation, but the constraints on the form
factors are nevertheless somewhat different than in BGL. 

While this \clnp fit is a simple modification of the CLN fit widely used by
BaBar and Belle, it still relies on heavy quark symmetry and model-dependent
input on subleading Isgur-Wise functions.  The reason is that both CLN and
\clnp use a cubic polynomial in $z = (\sqrt{w+1} - \sqrt2) / (\sqrt{w+1} +
\sqrt2)$ to parametrize the form factor $h_{A_1}$, with its four coefficients
determined by two parameters, $h_{A_1}(1)$ and $\rho_{D^*}^2$, derived from
unitarity constraints on the $B\to D$ form factor.  Therefore, we also 
consider a ``\nocln'' scenario, parametrizing $h_{A_1}$ by a quadratic
polynomial in $z$, with unconstrained coefficients,
\begin{equation}
h_{A_1}(w) = h_{A_1}(1) \big[ 1 - 8 \rho_{D^*}^2 z
  + (53.\, c_{D^*} - 15.) z^2 \big] , 
\end{equation}
keeping the same prefactors as in CLN, to permit comparison
between $\rho_{D^*}^2$ and $c_{D^*}$ (in the CLN fit $c_{D^*} =
\rho_{D^*}^2$).

The fit parameters in the BGL, CLN, \clnp, and \mbox{\nocln} fits are
summarized in Table~\ref{fitsummary}.  The results of these fits for
$|V_{cb}|$, $\rho^2_{D^*}$, $c_{D^*}$, $R_{1,2}(1)$, and $R_{1,2}'(1)$ are
shown in Table~\ref{tab:resVcbDs}.  The BGL, \clnp, and \nocln results are
consistent with each other, including the uncertainties, and the fit
quality.  The correlations of these four fit results for $|V_{cb}|$ are shown
in Table~\ref{tab:corrVcbDs} and have been derived by creating a
bootstrapped~\cite{Hayes:1988xc} ensemble of the unfolded distributions of
Ref.~\cite{Abdesselam:2017kjf}, using the published covariance. Each set of
generated decay  distributions in the ensemble is fitted with the BGL, CLN,
\clnp, and \mbox{\nocln} parametrizations, and the produced ensemble of
$|V_{cb}|$ values is used to estimate the covariance between them.  The
correlation of the CLN fit with either BGL, \clnp, or \nocln is substantially
below 100\%.  This reduces the tension between these fits to below $3\sigma$.

\begin{table}[t]\tabcolsep 2pt
\resizebox{\linewidth}{!}{
\begin{tabular}{c|cccc}
\hline\hline
& CLN & \clnp & \nocln & BGL \\
\hline
$|V_{cb}| \!\times\! 10^{3}$ &  $38.2 \pm 1.5$  & $41.5 \pm 1.9$ & $41.8 \pm 1.9$ & $41.5 \pm 1.8$ \\   
$\rho^2_{D^*}$	&  $1.17 \pm 0.15$ & $1.6 \pm 0.2$ & $1.8 \pm 0.4$  & $1.54 \pm 0.06$ \\
$c_{D^*}$	&  $\rho^2_{D^*}$  &  $\rho^2_{D^*}$  &  $2.4 \pm 1.6$  &
 fixed: 15./53. \\
$R_1(1)$	&  $1.39 \pm 0.09$ & $0.36 \pm 0.35$ & $0.48 \pm 0.48$  & $0.45 \pm 0.28$ \\
$R_2(1)$	&  $0.91 \pm 0.08$ & $1.10 \pm 0.19$ & $0.79 \pm 0.36$  & $1.00 \pm 0.18$  \\
$R_1'(1)$	&  fixed: $-0.12$  & $5.1 \pm 1.8$ & $4.3 \pm 2.6$  & $4.2 \pm 1.2$ \\
$R_2'(1)$	&  fixed: 0.11  & $-0.89 \pm 0.61$ & $0.25 \pm 1.3$  & $-0.53 \pm 0.42$ \\
\hline
$\chi^2$\,/ ndf	&  35.2 / 36  & 27.9 / 34 & 27.6 / 33 & 27.7 / 34 \\
\hline\hline
\end{tabular}
}
\caption{Summary of CLN, \clnp, \nocln, and BGL fit results.}
\label{tab:resVcbDs}
\end{table}

\begin{table}[t]
\begin{tabular}{l|cccc}
\hline\hline
   & $|V_{cb}|_{\rm CLN}$  &  $|V_{cb}|_{\rm \clnp}$  &
   $|V_{cb}|_{\rm \nocln}$  &  $|V_{cb}|_{\rm BGL}$ \\
   \hline
   $|V_{cb}|_{\rm CLN}$     & 1. & 0.75 & 0.69 & 0.76  \\
   $|V_{cb}|_{\rm \clnp}$   &    & 1.   & 0.95 & 0.97  \\
   $|V_{cb}|_{\rm \nocln}$  &    &      & 1.   & 0.97  \\
   $|V_{cb}|_{\rm BGL}$     &    &      &      & 1.    \\
\hline\hline
\end{tabular}
\caption{Correlation matrix of the four extracted $|V_{cb}|$ values.
For BGL the outer functions of Ref.~\cite{Grinstein:2017nlq} were used. All
results are derived by bootstrapping~\cite{Hayes:1988xc} the unfolded distributions of
Ref.~\cite{Abdesselam:2017kjf} using the published covariance.}
\label{tab:corrVcbDs}
\end{table}

\begin{figure*}[th]
\centerline{  
  \includegraphics[width=0.49\textwidth]{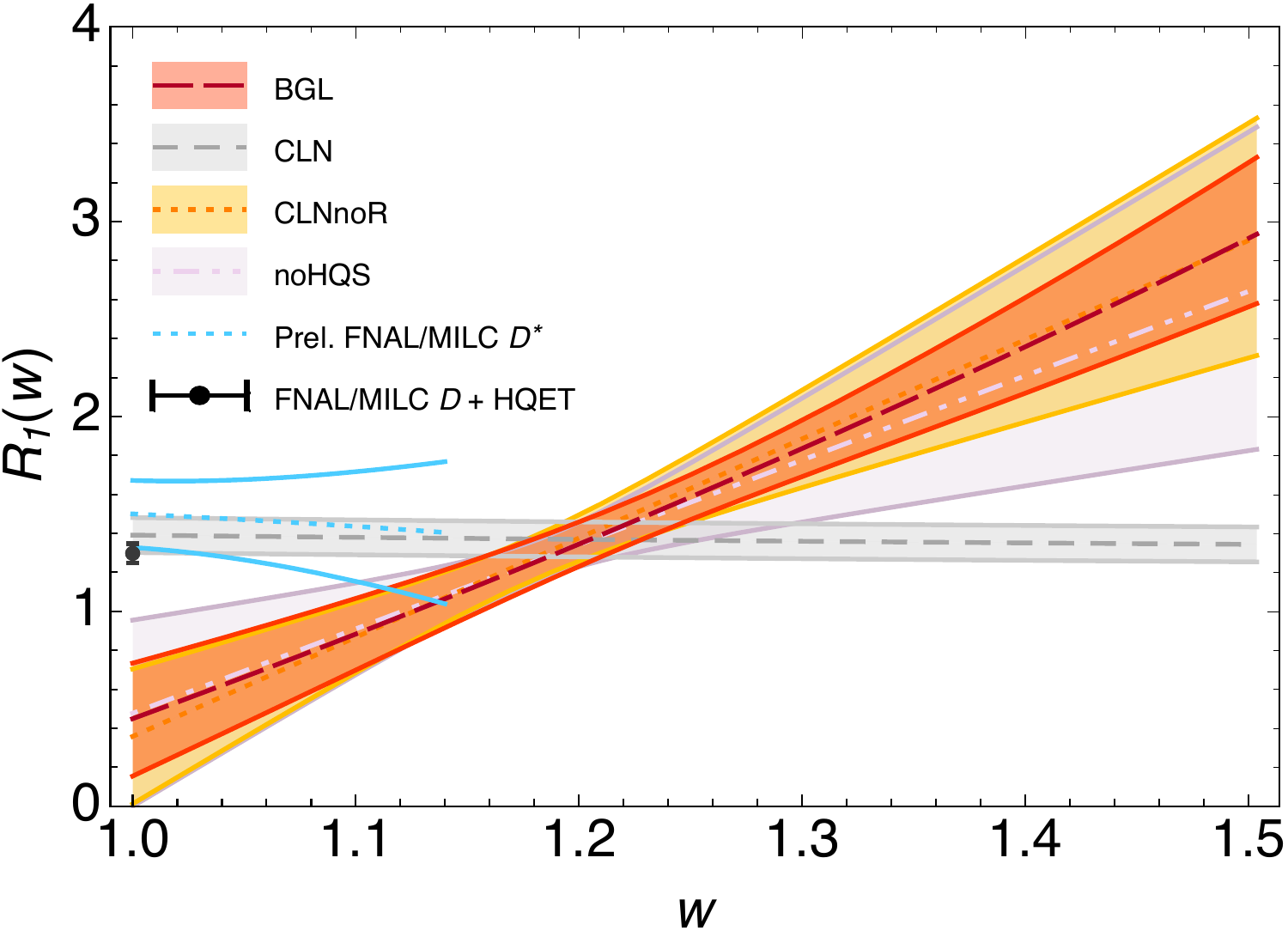} \hfil
  \includegraphics[width=0.51\textwidth]{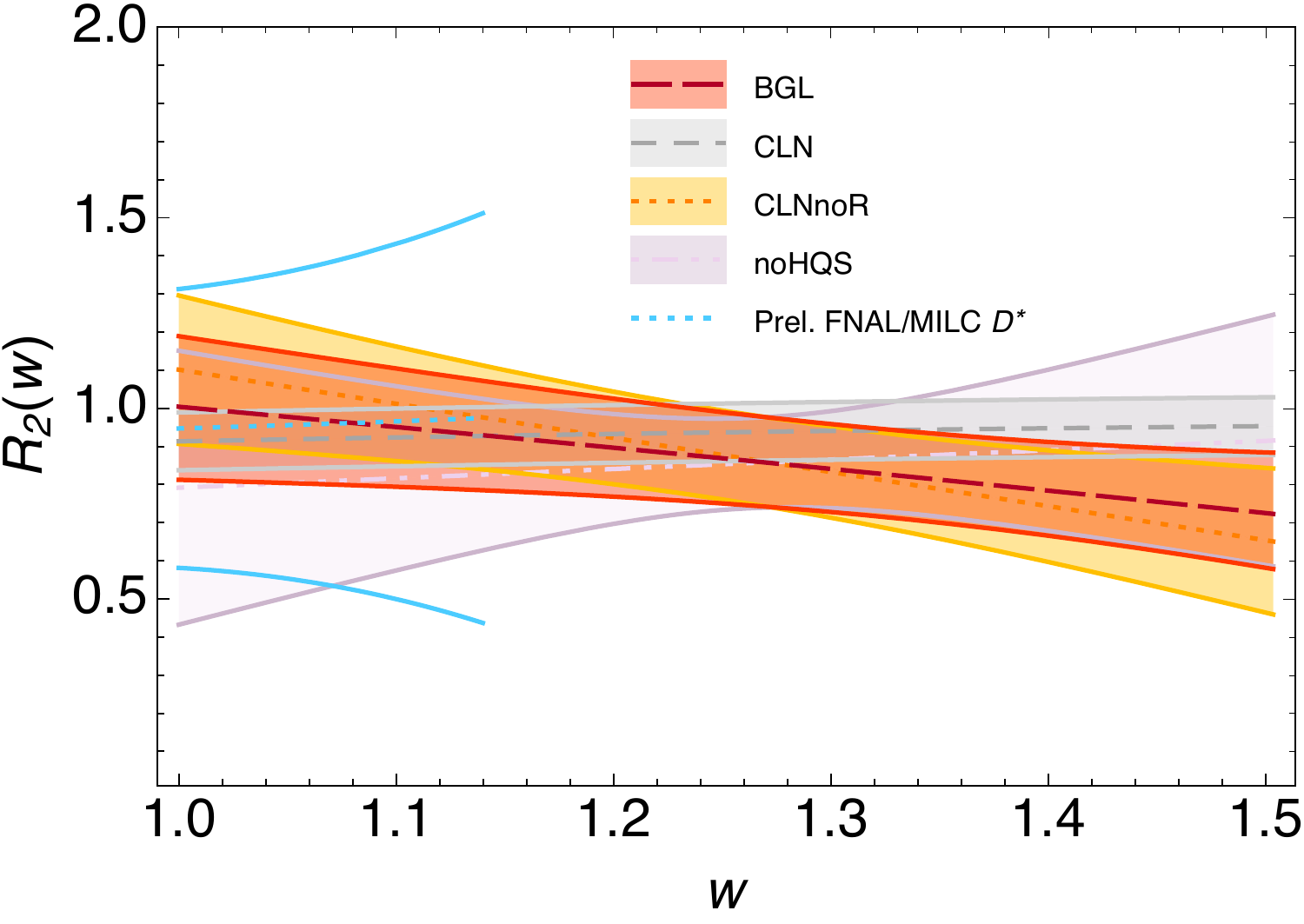}}
\caption{The form factor ratios $R_1(w)$ (left) and $R_2(w)$ (right) for the BGL
(red long dashed), CLN (gray dashed), \clnp (orange dotted) fits, and \nocln (purple dot-dot-dashed). The
BGL, \clnp, and \nocln fits for $R_1$ suggest a possibly large violation of
heavy quark symmetry, in conflict with lattice QCD predictions.  The blue
lines show our estimated bounds, based on preliminary FNAL/MILC lattice
results~\cite{BDsLatticeAllw}. The black data point for $R_1(1)$ follows from
the FNAL/MILC $B\to D\, \ell\, \bar \nu$ result and heavy quark symmetry (see
details in the text).}
\label{fig:R1_R2}
\end{figure*}

As soon as $R'_{1,2}(1)$ are not constrained to their values imposed in the
CLN framework, large deviations from those constraints are observed.  The
BGL, \clnp, and \nocln results favor a large value for $R_1'(1)$, in tension
with the heavy quark symmetry prediction, $R_1'(1) = {\cal
O}(\lqcd/m_{c,b},\, \alpha_s)$.

These aspects of the BGL, \clnp, and \nocln fits are also in tension with
lattice QCD results.  Recently the first preliminary lattice results were
made public on the $B\to D^*\ell\bar\nu$ form factors away from zero recoil,
at finite lattice spacing~\cite{BDsLatticeAllw}.  The results are fairly
stable over a range of lattice spacings.  Assuming that the continuum
extrapolation will not introduce a sizable shift (the chiral logs are not
large~\cite{Chow:1993hr, Randall:1993qg}) we can estimate the projections
for the $R_{1,2}(w)$ form factor ratios.  We approximate the predicted form
factors in a narrow range of $w$ using a linear form, with a normalization
and slope chosen such that they encompass all reported lattice points and
uncertainties in Ref.~\cite{BDsLatticeAllw}.  At zero recoil we obtain the
estimates $R_1(1) \simeq 1.5 \pm 0.2$ and $R_2(1) \simeq 0.95 \pm 0.45$,
which should be viewed as bounds on these values, as the actual lattice QCD
results will likely have smaller uncertainties.  Figure~\ref{fig:R1_R2} shows
$R_{1,2}(w)$ derived from the results of our fit scenarios, as well as
these lattice QCD constraints.

We can obtain another independent prediction for $R_1(1)$ based on lattice
QCD and heavy quark symmetry, using the result for the $B\to  D\, \ell
\bar\nu$ form factor~\cite{Lattice:2015rga}.  Using the
$\mathcal{O}(\lqcd/m_{c,b},\, \alpha_s)$
expressions~\cite{Bernlochner:2017jka}, the $f_+$ form factor (see Eq.~(2.1)
in Ref.~\cite{Lattice:2015rga}) and the subleading Isgur-Wise function $\eta$
are related at zero recoil via
\begin{align}
\frac{ 2 \sqrt{r_D} }{1+r_D} f_+(1) & =
  1 + \hat \alpha_s \bigg( C_{V_1} + C_{V_2}\, \frac{2r_D}{1+r_D}
    + C_{V_3}\, \frac{2}{1+r_D} \bigg) \nn\\
& \quad - (\varepsilon_c - \varepsilon_b)\, \frac{1-r_D}{1+r_D}\,
   [2\, \eta(1)-1] + \ldots ,\!\!\!\!
\end{align}
since other subleading Isgur-Wise functions enter suppressed by $w-1$. Here
$r_D = m_D/m_B$, $\varepsilon_{c,b} = \bar\Lambda/m_{c,b}$ is treated as in
Ref.~\cite{Bernlochner:2017jka}, and hereafter the ellipsis denotes
$\mathcal{O}(\varepsilon_{c,b}^2,\, \alpha_s\, \varepsilon_{c,b},\,
\alpha_s^2)$ higher order corrections.  Using $f_+(w = 1) = 1.199 \pm
0.010$~\cite{Lattice:2015rga} one finds $\eta(1) = 0.35 \pm  0.10$. The
uncertainty in this relation and the extracted value of $\eta(1)$ is
dominated by $\mathcal{O}(\lqcd^2/m_c^2)$ corrections parametrized by several
unknown matrix elements~\cite{Falk:1992wt}, which we estimate with
$\varepsilon_c^2 \sim0.05$.  Thus,
\begin{equation}\label{R11eq}
 R_1(1) = 1.34 - 0.12 \, \eta(1) + \ldots = 1.30 \pm 0.05 \,.
\end{equation}
(Recall that both the $\alpha_s$ terms and a
$\bar\Lambda/(2m_c)$ correction enhance $R_1(1)$.)
This estimate is shown with the black dot and error bar in the left plot in
Fig.~\ref{fig:R1_R2}.  It shows good consistency with our estimate from the
preliminary direct calculation of the $B\to D^*\ell\bar\nu$ form factors, as shown in the
region bounded by the blue curves.

Another clear way to see that the central values of the BGL, \clnp, and
\nocln fit results cannot be accommodated in HQET, without a
breakdown of the expansion, is by recalling~\cite{Bernlochner:2017jka} that
besides Eq.~(\ref{R11eq}), also
\begin{align}\label{R121}
R_2(1) &= 0.98 - 0.42\, \eta(1) - 0.54\, \hat\chi_2(1)
  + \ldots \,,\nn\\
R'_1(1) &= -0.15 + 0.06\, \eta(1) - 0.12\, \eta'(1) + \ldots\,, \\
R'_2(1) &= 0.01 - 0.54\, \hat\chi'_2(1)  + 0.21\,\eta(1) - 0.42\, \eta'(1)
  + \ldots\,.\nn
\end{align}
Here $\eta$ and
$\hat\chi_2$ are subleading Isgur-Wise functions.  Eqs.~(\ref{R11eq}) and
(\ref{R121}) have no solutions close to the BGL, \clnp, or \nocln fit
results in Table~\ref{tab:resVcbDs} with ${\cal O}(1)$ values for
$\eta(1)$,  $\eta'(1)$, $\hat\chi_2(1)$, and $\hat\chi'_2(1)$.

\begin{figure}[t]
\centerline{\includegraphics[width=\columnwidth]{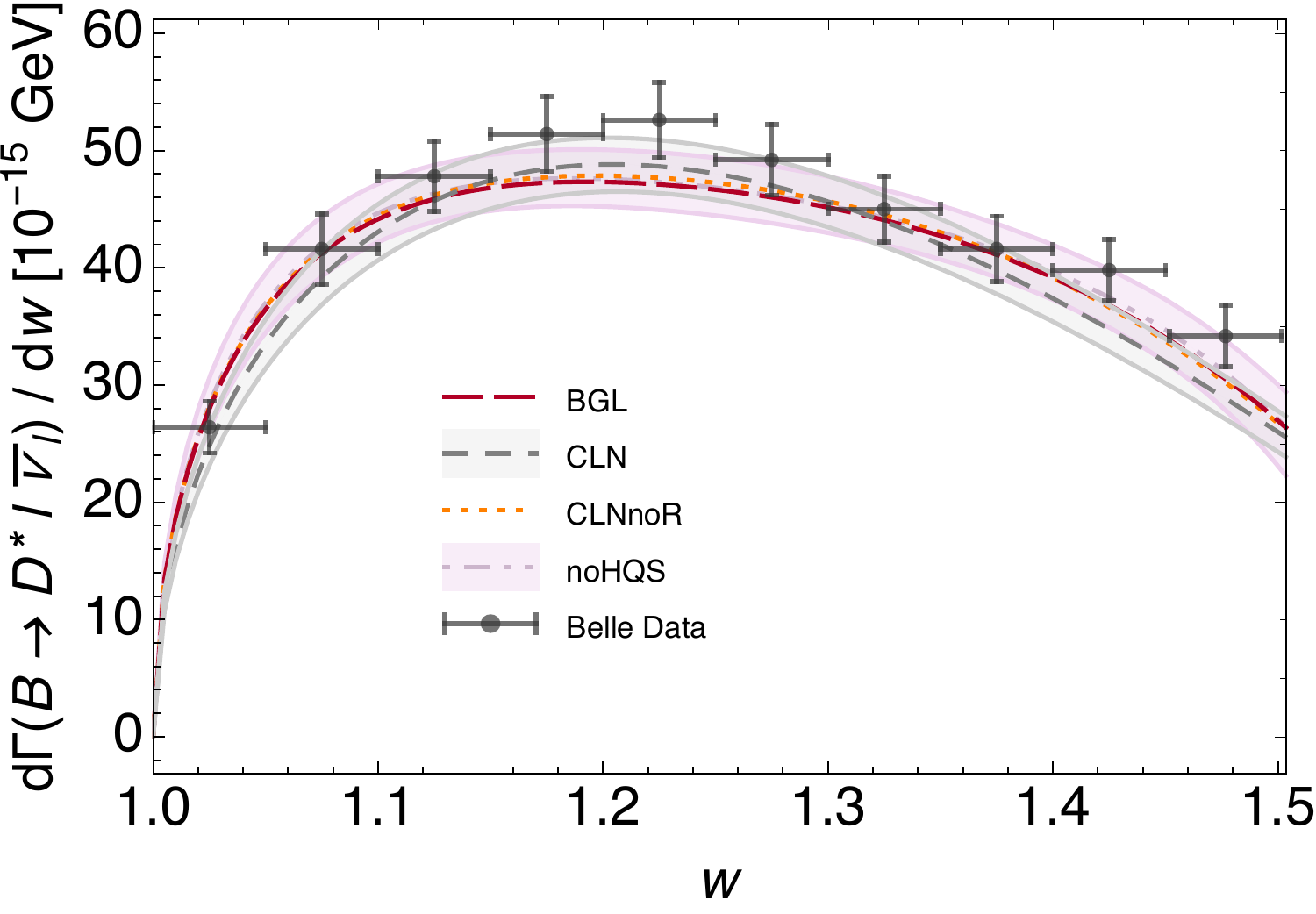}}
\caption{$\d\Gamma/\d w$ for the fit scenarios shown in
Fig.~\ref{fig:R1_R2}.}
\label{fig:dGdw}
\end{figure}

Figure~\ref{fig:dGdw} shows $\d\Gamma/\d w$ in the four fit scenarios, as
well as the Belle data~\cite{Abdesselam:2017kjf}.  The shaded bands show the
uncertainties of the CLN and \nocln fits, which are comparable to the
uncertainties of the other two fits.  The BGL, \clnp, and \nocln fits show
larger rates near zero and  maximal recoil, in comparison to CLN.  The CLN
fit shows a larger rate at intermediate values of $w$.

\section{Conclusions}

Our results show that the tensions concerning the exclusive and inclusive
determinations of $|V_{cb}|$ cannot be considered resolved.  The central
values of the BGL, \clnp, and \nocln fits, which all give good descriptions
of the data, suggest possibly large deviations from heavy quark symmetry. 
These results are also in tension with preliminary lattice QCD predictions
for the form factor ratio $R_1$, which use the same techniques as for the
determination of ${\cal F}(1)$ used to extract $|V_{cb}|$ from $B\to
D^*\ell\bar\nu$.  If the resolution of the tension between lattice QCD and
the fits for $R_1$ is a fluctuation in the data, then we would expect the
extracted value of $|V_{cb}|$ to change in the future.  If the resolution of
the tension is on the lattice QCD side, then it may also affect the
calculation of ${\cal F}(1)$ used to extract $|V_{cb}|$.  We look forward to
higher statistics measurements in the future, and a better understanding of
the composition of the inclusive semileptonic rate as a sum of exclusive
channels~\cite{Bernlochner:2012bc, Bernlochner:2014dca}, which should
ultimately allow unambiguous resolution of these questions.

\acknowledgments

FB and ZL thank Prof.\ Toru Iijima for organizing the ``Mini-workshop on 
$D^{(*)}\tau\nu$ and related topics", and the kind hospitality in Nagoya,
where this work started. 
We also thank the Aspen Center of Physics, supported by the NSF grant 
PHY-1066293, where this paper was completed. 
We thank Ben Grinstein and Bob Kowalewski for helpful conversations, not only
over sushi and sake.
FB was supported by the DFG Emmy-Noether Grant No.\ BE~6075/1-1. 
ZL and MP were supported in part by the U.S.\ Department of Energy under
contract DE-AC02-05CH11231. 
DR acknowledges support from the University of Cincinnati.

\bibliographystyle{apsrev4-1}
\bibliography{VcbComp}

\end{document}